# Metalloporphyrins on oxygen-passivated iron: Conformation and order beyond the first layer

David Maximilian Janas[1,*], Andreas Windischbacher[2], Mira Sophie Arndt[1], Michael Gutnikov[1], Lasse Sternemann[1], David Gutnikov[1], Till Willershausen[1], Jonah Elias Nitschke[1], Karl Schiller[1], Daniel Baranowski[3], Vitaliy Feyer[3], Iulia Cojocariu[3,4], Khush Dave[5], Peter Puschnig[2], Matija Stupar[1], Stefano Ponzoni[1], Mirko Cinchetti[1] and Giovanni Zamborlini[1,#] .

[1]*TU Dortmund University, Department of Physics, 44227 Dortmund, Germany.*

[2]*Karl-Franzens-Universität Graz, Institut für Physik, NAWI Graz, 8010 Graz, Austria.*

[3]*Peter Grünberg Institute (PGI-6), Forschungszentrum Jülich GmbH, 52428 Jülich, Germany.*

[4]*Dipartimento di Fisica, Università degli Studi di Trieste, via A. Valerio 2, 34127 Trieste, Italy.*

[5]*Department of Physics, Indian Institute of Science Education and Research Bhopal, Bhopal, MP 462066, India.*

Corresponding authors: * david.janas@tu-dortmund.de; # giovanni.zamborlini@tu-dortmund.de.

**Abstract:**
On-surface metal porphyrins can undergo electronic and conformational changes that play a crucial role in determining the chemical reactivity of the molecular layer. Therefore, accessing those properties is pivotal for their implementation in organic-based devices. Here, by means of photoemission orbital tomography supported by density functional theory calculations, we investigate the electronic and geometrical structure of two metallated tetraphenyl porphyrins (MTPPs), namely ZnTPP and NiTPP, adsorbed on the oxygen-passivated Fe(100)-$p$(1 × 1)O surface. Both molecules weakly interact with the surface as no charge transfer is observed. In the case of ZnTPP our data correspond to those of moderately distorted molecules whereas NiTPP exhibits a severe saddle-shape deformation. From additional experiments on NiTPP multilayer films, we conclude that this distortion is a consequence of the interaction with the substrate, as the NiTPP macrocycle of the second layer turns out to be flat. We further find that distortions in the MTPP macrocycle are accompanied by an increasing energy gap between the highest occupied molecular orbitals (HOMO and HOMO-1). Our results demonstrate that photoemission orbital tomography can simultaneously probe the energy level alignment, the azimuthal orientation, and the adsorption geometry of complex aromatic molecules even in the multilayer regime.

**Introduction:**

Tetrapyrrolic compounds, such as porphyrins and phthalocyanines, have raised a lot of attention due to their remarkable chemical stability and outstanding tunability.[1,2] Porphyrins, in particular, play a key role in several biological and catalytic processes[2,3] and are employed as molecular building blocks in several technological applications, *e.g.* chemical sensors,[4] solar cells[5] and OLEDs.[6]

Their chemical and physical properties can be controlled on-demand either by attaching different peripheral substituents to their tetrapyrrolic unit, or *via* the incorporation of different transition metal ions in their center. On proper surface templates, porphyrins can be confined in two-dimensional (2D) arrays, where the chemical state and the adsorption properties of the molecules are well-defined and homogeneous.[2,7] This opens the possibility of utilizing them in a variety of electronic and spintronic devices[8,9] and exploiting them as single-atom catalysts (SACs).[10]

In fact, these supramolecular structures act as a network, stabilizing ordered arrays of single metal atom active sites. The chelated ion center becomes available for the axial coordination of external ligands[11–15] and for the chemical conversion of small molecules.[3,16,17] In this regard, the oxidation state of the chelated metal ion, which can be strongly influenced by the molecule-substrate interaction, is of utmost importance for determining the reactivity of the porphyrin layer.[13,18] Indeed, when porphyrins are deposited atop Cu and Ag, charge transfer from the substrate to the molecular layer takes place, leading to the filling of the lowest unoccupied molecular orbitals (LUMOs),[14,19] followed by the reduction of the chelated metal ion.[20–22] In parallel, the conformation of the macrocycle can induce variations in the electronic structure,[23] altering not only the optical properties[24] but affecting also the capability of the metal center of anchoring ligands,[25] which would be the first step for a catalytic reaction. For example, it has been demonstrated that the macrocycle saddling (*i.e.* tilting of the macrocyclic pyrrole rings) can promote the coordination of CO in a *cis-* or *trans-*configuration either to the pyrrole groups or the metal center,[26] respectively, while the ligation does not occur for a completely planar macrocycle.[27]

In this context, passivating the metal surface with oxygen plays a key role, as it significantly influences both the electronic and the structural properties of the on-surface porphyrin. On the one hand, oxygen passivation helps to preserve the electronic properties of the molecular adsorbate by reducing the molecule-substrate interaction,[13,28] which is rather strong in the case of ferromagnetic substrates.[29] On the other hand, it has a profound impact on the molecular adsorption geometry, causing distortions in the molecular backbone.[30] For this reason, we investigated the electronic and geometric structure of two different metal-containing porphyrins, namely zinc and nickel tetraphenyl porphyrins (ZnTPP and NiTPP, respectively) atop the oxygen-passivated Fe(100)-$p(1 \times 1)$O surface,[31] for simplicity referred in the following as Fe-O.

While the Zn(II)TPP has a fully closed shell ($3d^{10}$ electron configuration), suggesting a rather inert character of the metal center, the Ni(II)TPP has a $3d^8$ configuration, with a completely unfilled orbital level ($d_{x^2-y^2}$). Upon deposition on the surface template, this may lead to differences in both the adsorption configuration and electronic structure of the molecular film. Recent studies investigating the monolayer and multilayer of ZnTPP[30,32,33] and NiTPP[30,34,35] atop Fe-O were able to shed light on the molecular adsorption geometries and give a first insight into the interfacial electronic structure. The NiTPP molecules in direct contact with the substrate self-assemble in a commensurate (5 x 5)R37° superstructure and their macrocycle undergoes a severe saddling. Instead, the data reported for the ZnTPP indicate that the molecules arrange in a (5 × 5) structure, with their tetrapyrrolic backbone distorted in a similar, but less prominent saddle shape conformation.[30] Regarding the electronic structure, ultraviolet photoemission spectroscopy (UPS) reveals that, upon adsorption atop the passivated Fe surface, the electronic structure of both NiTPP and ZnTPP in the mono- and multilayer regimes are barely affected, leaving the molecules in a quasi-free state, where the initial oxidation state is preserved.[32,36]

One drawback of UPS applied to molecule/metal interfaces is that the resulting spectra often consist of several broadened molecular features that partially or fully overlap in energy. This complicates the determination of the orbital origin for each peak, which is vital for a deeper understanding of the substrate-molecule interaction, as well as intermolecular interactions for multilayer films.[37] However, with specific assumptions about the photoemission matrix element, it is possible to compare the angular (momentum) distribution of the photoemitted electrons to the modulus square of the Fourier transform $|FT|^2$ of a specific molecular orbital, which can be calculated by means of density functional theory (DFT). This approach is called photoemission orbital tomography (POT) and it not only enables to accurately identify individual molecular orbitals but also to reconstruct the related charge distribution.[37–40] Based on this, POT has proven to be a valuable tool for investigating organic mono- and bilayers systems on surfaces, giving access to the electronic structure and also the molecular adsorption geometry.[41,42] Moreover, it was shown just lately that this technique is also able to detect geometric distortions of aromatic complexes with extremely high accuracy.[41,43]

In this study, we employ photoemission orbital tomography to study the frontier orbital structure of both monolayer ZnTPP and NiTPP adsorbed on Fe-O, providing further information regarding their energy level alignment. At the same time, we can access the adsorption conformation of the porphyrin macrocycle quantifying its saddling degree. Furthermore, we investigate the growth of a second layer of NiTPP: while it produces the same low energy electron diffraction (LEED) pattern as the monolayer, the NiTPP macrocycle does not undergo any geometrical distortion. This stands in contrast to the first layer, where the molecule takes a pronounced saddle-shape adsorption configuration.

**Methods:**

All the experiments were performed under ultra-high vacuum conditions with a base pressure of $1\cdot10^{-10}$ mbar. A 400 nm thick Fe film was deposited on a clean MgO(100) substrate, which is initially cleaned by cycles of $Ar^+$ sputtering and annealing at 870 K. Following the recipe of Picone et al.[44] the Fe-O surface was obtained by exposing the clean Fe(100) sample to 30 L of $O_2$ (5 min at $1.3\cdot10^{-7}$ mbar) while the sample was kept at 820 K. Afterwards the sample was flash annealed to 870 K to remove the excess of oxygen from the surface.[44] For every new preparation, the Fe sample was cleaned by cycles of gentle sputtering ($Ar^+$ ions with 0.5 keV kinetic energy) and subsequent annealing at 870 K for 5 min. Cleanliness and surface quality were checked after each preparation by momentum-resolved photoemission spectroscopy (see Supplementary S1).

The deposition rates for the monolayer ZnTPP and NiTPP were first estimated using a quartz crystal microbalance and then calibrated by stepwise deposition and subsequent monitoring of the emerging (5 x 5) and (5 x 5)R37° LEED patterns, which are well-known from literature.[30] The rates for the growth of one full saturated monolayer (1 ML) were estimated as 0.1 ML/min for both molecules with a deposition temperature of 543 K and 531 K for ZnTPP and NiTPP, respectively. For the multilayer experiments on NiTPP the rate was increased to 0.4 ML/min to match approximately the rate from Ref.[34]

The momentum-resolved photoemission data on ZnTPP were acquired at room temperature using a KREIOS PEEM (Specs GmbH) operating in momentum mode.[45] Such an instrument allows to record 2D momentum maps with an approximate momentum field-of-view of $k_x, k_y \in [-2.0,+2.0]$ Å$^{-1}$ at a constant kinetic energy. The PEEM is coupled to a laser based light source, which relies on high harmonic generation to provide *p*-polarized fs-XUV pulses with a photon energy of 29.7eV (25th harmonic of the driving laser frequency). The present microscope settings led to an energy resolution < 200 meV.

The system is driven by a commercial light conversion Ytterbium laser with 1030 nm central wavelength, 242 fs pulse duration, 200 kHz repetition rate and 50 W average power.

Photoemission experiments on the NiTPP films were performed at the NanoESCA beamline at the Elettra synchrotron in Trieste, Italy, using a NanoESCA PEEM (Focus GmbH). Here, *p*-polarized light with an energy of 40 eV was used while the sample was cooled to 100 K. The energy resolution was <100 meV.

The experimental data are complemented by DFT calculations of the molecules in different molecular conformations on a periodic, oxygen-covered Fe slab, as well as in the gas phase. The periodic calculations are performed with the Vienna Ab Initio Simulation Package (VASP) version 5.4.4,[46,47] to obtain the projected density of states for the composite interfaces. The arrangements of the molecules in the unit cell were taken from a previous STM study of Fratesi et al.[30] and molecules were visualized using the VESTA software.[48] Exchange-correlation effects were approximated by the functional of Perdew-Burke-Ernzerhof (PBE).[49] Additionally, we included an effective Hubbard-U correction[50] to the metal *d*-states $U_{eff}$=3eV to improve the treatment of on-site correlations at the metal center of the porphyrin complexes. Van der Waals contributions were treated with Grimme's D3 dispersion correction.[51] We used the projector-augmented wave (PAW) method[52] assuming an energy cutoff of 400 eV. All calculations were spin-polarized and the ionic positions optimized until forces were below 0.01 eV/Å. The interfaces were simulated within the repeated slab approach using six substrate layers, two of which were held fix during optimization, and a 25 Å vacuum layer between the slabs. To avoid spurious electrical fields, a dipole layer was inserted in the vacuum region.[53] The Brillouin zone was sampled on a Gamma-centered grid of 3x3x1 k-points.

For our POT analysis, we extended our study with gas phase calculations using the quantum chemistry package ORCA version 5.0.1.[54,55] Reducing the size of the systems seems reasonable, as the molecules are effectively electronically decoupled from the substrate. Furthermore, it enables the use of the more sophisticated hybrid-functional (B3LYP[56,57]) to describe the electronic structure of the porphyrins. The def2-TZVP basis set[58] was employed together with Grimme's D3 dispersion correction.[51] From the resulting Kohn-Sham orbitals, momentum maps are simulated by taking isokinetic cuts of the Fourier-transformed orbitals via the kMap software package.[39]

## Results and Discussion:

Monolayer ZnTPP on Fe(100)-p(1 × 1)O

Fig. 1 shows the valence band spectrum, integrated in momentum space over a region of ± 1.8 Å$^{-1}$, for both the bare Fe-O and 1 ML of ZnTPP deposited on Fe-O. These two energy distribution curves (EDCs) are normalized at the Fermi edge where the photoemission intensity can be solely attributed to the substrate and no molecular features are expected.[59] By comparing the bare Fe-O and the ZnTPP/Fe-O spectra, we observe that two new features at binding energies (BEs) of 1.5 eV and 3.5 eV arise upon molecular adsorption. While the latter was previously assigned to the phenyl moiety, the former originates from the tetrapyrrole macrocycle and exhibits a double-peak structure.[33,59] To investigate the origin of this splitting, we measured the energy region up to BE = 2.1 eV with increased integration time yielding the spectrum displayed in the right panel Fig. 1. We evaluate the contribution of the two underlying peaks by fitting the section from 0.8 eV to 2.3 eV BE using two Gaussian functions with a linear background. The corresponding Gaussians are centered at (1.3 ± 0.1) eV and (1.6 ± 0.1) eV with FWHMs of (0.3 ± 0.1) eV and (0.4 ± 0.1) eV, respectively. Their corresponding curves are depicted below the EDCs in the right panel of Fig 1.

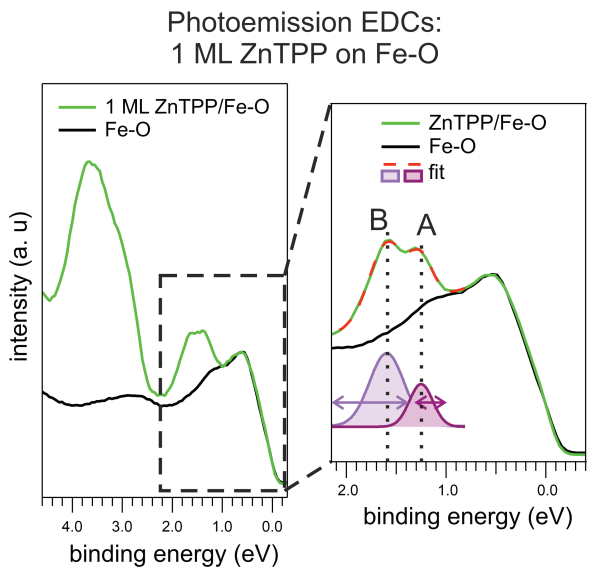

Figure 1: Photoemission data of 1 ML ZnTPP on Fe-O. The left panel shows the momentum-integrated spectra of the pristine Fe-O surface (black curve) and the Fe-O covered with 1 ML ZnTPP (green curve). The panel on the right displays a measurement with increased integration time around the first prominent molecular feature at 1.5eV BE. The double-peak structure of this feature is fitted by two Gaussians and a linear background. The two purple arrows indicate the energy regions that were used to generate averaged momentum maps, which are displayed in Fig. 2.

To assign these peaks to a specific molecular orbital, we analyze their intensity distribution as a function of the momentum coordinates. For this purpose, we averaged the momentum maps over energy ranges that are indicated by the arrows in Fig. 1. Selecting these ranges ensures that the different orbital contributions do not mix up in the k-resolved maps. Subsequently, the maps were symmetrized according to the D$_4$ symmetry of the substrate, resulting in the patterns presented in the upper parts of the panels in Fig. 2c and 2f. The latter step compensates for intensity variations due to the polarization factor and enables a one-to-one comparison of the processed experimental maps to simulated momentum distributions – the cornerstone of the POT framework.[37,39] Comparing the experimental to the simulated momentum maps allows us to assign unambiguously each momentum pattern to a distinct molecular orbital. The symmetrizing process of the theoretical maps is further illustrated in Fig. S10 (SI, Sec. S5). Moreover, since photoelectron spectroscopy averages over large surface areas, our data encode information about the geometrical structure of the molecules in the adsorbed monolayer, which we are going to unveil in the following.

At first, we compare our experimental data with the simulated maps obtained from a ZnTPP molecule with a planar macrocycle moiety, as depicted in Fig. 2a. We start from this specific conformation of the backbone since it has already been proposed for other on-surface porphyrins.[19,60] Hereby, we recognize that the two discerned peaks can be ascribed to the HOMO and HOMO-1 of gas phase ZnTPP. The real space electron distributions of these orbitals are pictured in Fig. 2b. The comparison between the ensuing momentum patterns and the experimental maps originating from peak A and B is displayed in Fig. 2c (bottom and top halves, respectively). By considering the $D_4$ symmetry domains on the Fe-O surface, we can further conclude that ZnTPP arranges with its N-Zn-N axis aligned at an angle of 17° with respect to the [001] substrate direction (see Fig. S2 in the SI for further details).

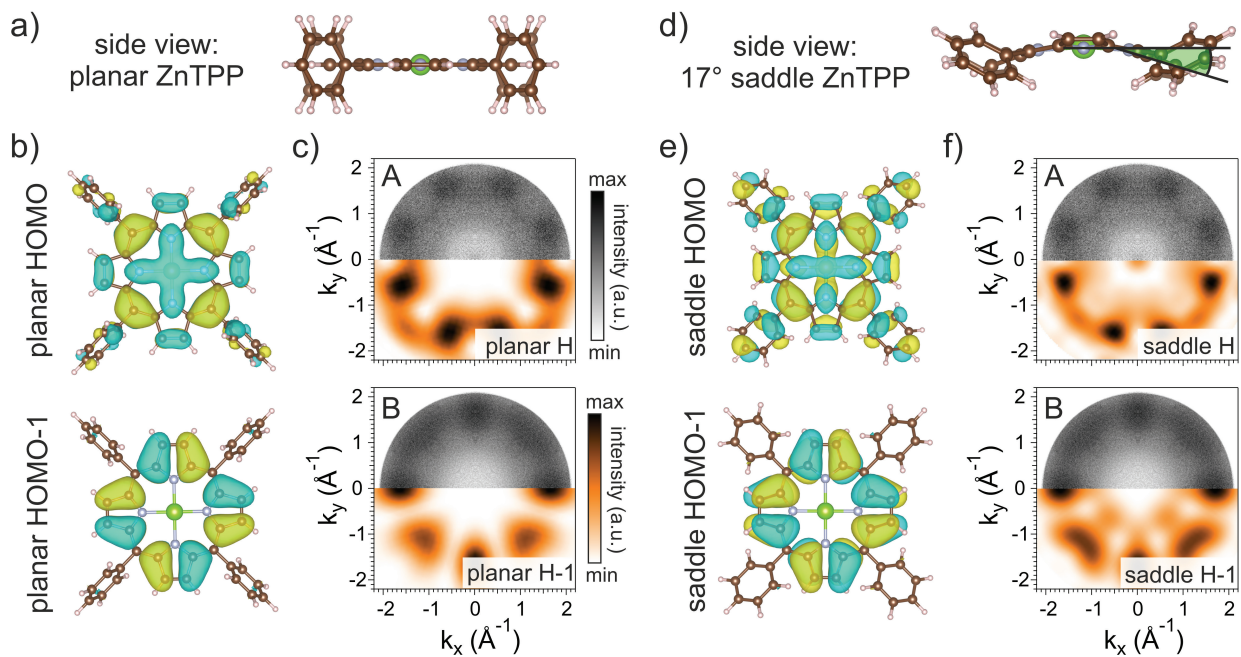

*Figure 2: Photoemission maps of 1 ML ZnTPP on Fe-O. In a) the ball-and-stick model of a planar ZnTPP molecule is displayed, which is complemented in b) by the corresponding real space orbitals of HOMO and HOMO-1. The simulated momentum patterns of these orbitals in a planar conformation are shown in the bottom regions of the panels in c), where they are compared to the symmetrized experimental maps belonging to peak A and B. In d) the model of a 17° saddle-shaped ZnTPP molecule is shown. The corresponding real space distributions of HOMO and HOMO-1 for this optimized saddle geometry are displayed in e). A comparison of the associated simulated momentum maps and the experimental maps is depicted in f).*

However, subtle but notable features, such as the faint intensity close to the center of the map of peak A, as well as the elongation of the main lobes at higher momenta, can only be reproduced when a saddle-shape conformation of the molecule is taken into account in our simulations of the $|FT|^2$ (an overview of maps from different conformations is presented in Fig. S5). We define the saddling angle as the average angle enclosed by the plane of the macrocycle and the plane of the pyrrole units, as indicated in the ball-and-stick model in Fig. 2d. Such a saddle-shaped adsorption geometry, with the ZnTPP molecule slightly rotated off the substrate axis, is in good agreement with the adsorption geometry determined by the combined STM and DFT study of Fratesi *et al.*[30]

Looking at Fig. S5, we notice that by increasing the saddling degree the overall appearance of the momentum maps changes. This is a consequence of the fact that structural changes in the molecular complex alter the shape of the orbitals in real space, which likewise affects their appearance in momentum space. Therefore POT should be able to also quantify the deformation of the molecules, as recently suggested for polycyclic aromatic compounds.[43] To determine the degree of distortion, we perform additional periodic calculations for the molecule on the Fe-O surface and relax its adsorption geometry. The real space distributions of HOMO and HOMO-1 resulting from the saddle-shape structure are depicted in Fig. 2e. The corresponding |FTs|² are displayed in Fig. 2f right below the experimental maps, exhibiting excellent qualitative agreement. For the optimized geometry, we find a saddling angle of 17°, as presented in Fig. 2d. In case of the HOMO-1, where the charge is located solely on the carbon atoms, this saddling does not induce any notable changes in the electron density aside from an inevitable tilting of the orbital lobes due to a tilting of the pyrrole units. The HOMO instead, which has a noticeable charge contribution from the nitrogen and central metal ion $p_z$ states, appears to be more sensitive to structural modifications. Particularly at the porphyrin center, the saddling leads to evident variations in the electron density and our calculations reveal that these variations are accompanied by a shift of the HOMO towards lower BE. Consequently, the projected density of states (PDOS) of our ZnTPP/Fe-O calculations (see Fig. S7a in the SI) exhibits a double-peak structure due to well-separated HOMO and HOMO-1 levels, coinciding well with the UPS spectrum in Fig. 1. The theoretically derived energy splitting of 370 meV, matches the experimentally obtained value of (300 ± 100) meV.

NiTPP on Fe(100)-*p*(1 × 1)O from monolayer to multilayer

As stated in the introduction, different transition metal ions can be embedded inside the porphyrin offering the possibility of tuning the chemical, electronic and transport properties of the interface. For this reason, we move now from ZnTPP to NiTPP and study its interaction with the Fe-O surface. Analogous to ZnTPP/Fe-O, we start by discussing the momentum-integrated valence band measured for the 1 ML NiTPP film on top of Fe-O. The corresponding spectrum is shown in Fig. 3a, together with the one of pristine Fe-O. In contrast to ZnTPP, the first molecular feature at 1.5 eV BE does only exhibit a single peak with a Gaussian FWHM of (0.5 ± 0.1) eV. This value is slightly larger than the two individual FWHMs obtained for ZnTPP.

Notably, the HOMO and HOMO-1 do not contain any *d* states of the central metal ion and their shapes are almost identical for ZnTPP and NiTPP. However, the absence of a clear double peak structure suggests that in the case of NiTPP on Fe-O, the HOMO and HOMO-1 are almost degenerate, similarly to the NiTPP/Cu(100) interface.[19] To improve the signal-to-noise ratio, we obtain the corresponding momentum map by summing over all the maps in the energy range of the FWHM (indicated by the red arrow in Fig. 3a). It is worth noticing that due to the different photon energy (with respect to the ZnTPP experiments), the contribution from the substrate is no longer negligible, requiring an additional background removal procedure (further details on this procedure are available in the SI Sec. S5). The final symmetrized and background-corrected map is displayed in Fig. 3b.

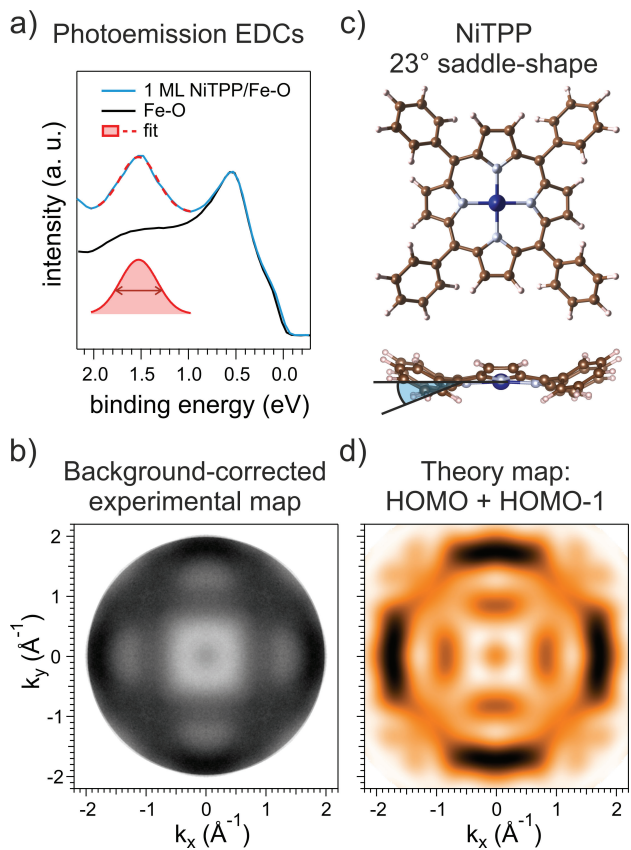

*Figure 3: Photoemission data of the frontier peak of 1ML NiTPP on Fe-O. a) Close up of the EDCs alongside the Gaussian determined from fitting the first molecular feature. b) Experimental photoemission map of the 1ML NiTPP film integrated over the energies, which are indicated by the red arrow in a). c) Ball-and-stick model of a NiTPP molecule with strong saddle-shape conformation. In d) the simulated photoemission map for the sum of HOMO and HOMO-1 for a 23° saddle-shape NiTPP is displayed.*

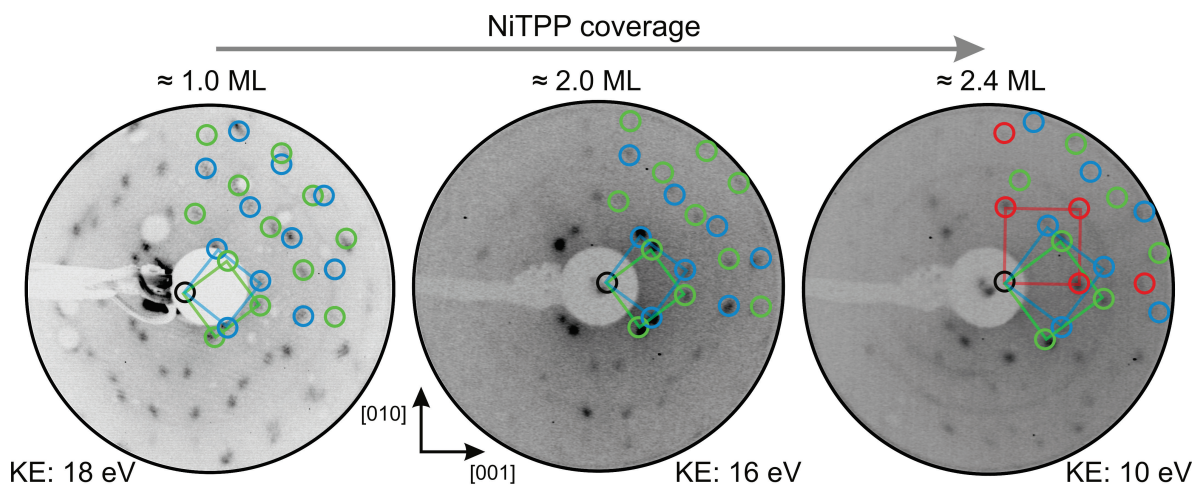

*Figure 4: LEED patterns of different coverages of NiTPP adsorbed on Fe-O. From left to right coverages of 1 ML, 2 ML and 2.4 ML are shown. The diffraction spots of the (5 × 5)R37° superstructure, which gives rise to two symmetry-equivalent domains, are indicated by the blue and green circles together with the corresponding unit cells. For the 2.4 ML coverage, an additional pattern corresponding to a (5 × 5) superstructure (red circles) alongside its unit cell (red square) are displayed.*

Next up, we employ an analogous POT analysis as for ZnTPP, with the only difference that we treat the frontier orbitals as degenerate and compare our photoemission data to the averaged $|FTs|^2$ of HOMO and HOMO-1. An overview of the resulting maps as a function of the conformation can be found in the SI, Fig. S6. Consistent with literature,[30] the best visual agreement in the comparison is achieved when the molecules take an angle of 35° between the molecular N-Ni-N axis and the substrate [001] axis (see Fig. S2). However, contrary to ZnTPP, the simulated pattern of planar NiTPP fails to reproduce fully our experimental data, making necessary the introduction of a significant macrocycle deformation in order to capture all the prominent features adequately.

From periodic calculations for NiTPP atop Fe-O, we derive the molecular structure shown in Fig. 3c, which embodies a saddling angle of 23°. The $|FT|^2$ corresponding to that saddle distortion (see Fig. 3d) is in accord with our photoemission measurements. Furthermore, the PDOS resulting from our periodic calculations (see Fig. S7b in the SI) resembles closely the measured EDC in Fig. 3a. According to theory, 220 meV separate the HOMO and HOMO-1. However, this splitting is not large enough to produce distinct peaks; instead, the two contributions merge into one broad feature, resulting in the enhanced FWHM we observe experimentally.

The fact that NiTPP adsorbs in a severely distorted configuration is in line with the findings of previous reports stating a more pronounced saddle-shape conformation for NiTPP in comparison to ZnTPP.[30] Notably, while our DFT calculations indicate that deformations of the macrocycle are accompanied by an enhanced lifting of the HOMO and HOMO-1 degeneracy, DFT also shows that the resulting energy splitting $\Delta E$ is generally larger in ZnTPP than in NiTPP and therefore depends strongly on the choice of the chelated metal ion. Already in an optimized planar gas phase conformation, these two orbitals are shifted by 170 meV for ZnTPP but only 60 meV for NiTPP. The same trend is reflected by the optimized saddle-shape geometry, as $\Delta E = 370$ meV for ZnTPP decreases to $\Delta E = 220$ meV for NiTPP. Noteworthy, the increased splitting is mostly related to a shift of the HOMO towards lower BEs, while the BE of the HOMO-1 remains rather unaffected.

Next, we investigate the geometrical and electronic properties of a NiTTP multilayer film and explore if any changes arise from the inevitably reduced molecule-substrate interaction. We start by reporting the LEED patterns of the NiTPP/Fe-O at increasing coverages, namely 1, 2 and 2.4 ML. The corresponding images are shown in Fig. 4. Notably, the pattern of the 2ML film shows very sharp spots that coincide with the ones reported for a single ML,[30] exhibiting the same (5 x 5)R37° superstructure. In contrast, increasing the coverage even more leads to the emergence of additional spots alongside a minor blur of the overall pattern. The additional spots match a (5 x 5) overlayer structure. In the presented LEED images, the symmetry-equivalent unit cells of the (5 x 5)R37° superstructure are superimposed as green and blue squares and the unit cell of the (5 x 5) superstructure is indicated by a red square. These findings point towards an ordered growth of the NiTPP molecules even beyond the first layer mediated by non-negligible intermolecular interactions.

Moving now to the electronic structure of the interface beyond the monolayer coverage, in Fig. 5a we compare the EDCs of the clean Fe-O substrate, the 1 ML and the 2 ML films of NiTPP/Fe-O. Upon going from 1 ML to 2 ML, the most notable differences induced by the second layer are the emergence of an additional peak at 2.6 eV and a shift of the first peak from 1.5 eV towards higher binding energies.

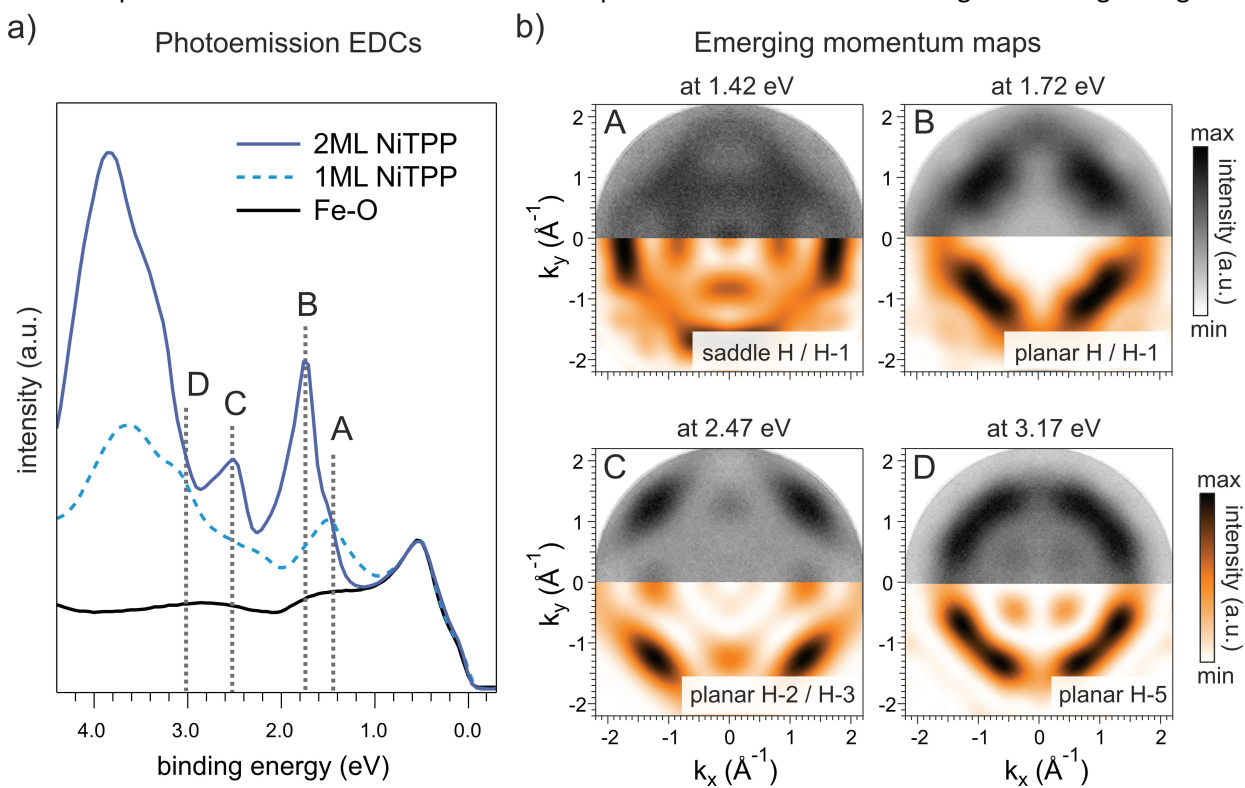

Figure 3: Photoemission data for the 2ML NiTPP film. In a) the momentum-integrated photoemission spectra for the Fe-O substrate as well as the 1 ML and 2 ML NiTPP films are shown. The dotted vertical lines labelled A,B,C, and D indicate energies from where selected momentum maps were extracted. In b) the corresponding momentum maps are displayed in the top parts of each panel with the respective simulated theory maps right below them.

Furthermore, a kink at around 1.6 eV in the 2 ML spectrum indicates that this peak consists of a main peak and an additional shoulder at slightly lower binding energies. The most prominent features that have a clear signature in momentum are highlighted in the EDC in Fig. 5a, and the corresponding momentum maps are lined up in the upper halves of the panels in Fig. 5b. The fact that these maps show well-defined features suggests a uniform azimuthal orientation also for the second NiTPP layer enabling us to perform a POT analysis also for the 2 ML NiTPP/Fe-O system, which we will discuss in more detail in the following part.

For the map that originates from the shoulder at 1.4 eV (map A in Fig. 5b), we observe a strong resemblance with the HOMO/HOMO-1 map of the saddle-shape distorted NiTPP that we previously reported in Fig. 3b showing the same characteristic features around the center of the map. Since the energy position of this shoulder matches very precisely the peak position of the 1 ML NiTPP film, we infer that this signal stems from the first molecular layer. This implies that even when subsequent molecular layers are deposited, the molecules in direct contact with the substrate retain their distorted configuration. Strikingly, also peak B at around 1.7 eV originates from the HOMO/HOMO-1 levels, however, the corresponding momentum map (map B) shows no sign of the previously observed distortion-related features. Instead, a comparison with simulated momentum distributions for planar NiTPP reveals that peak B and the two other ones (C, D in Fig. 5a) can be assigned to the gas phase orbitals of molecules with a flat macrocycle. Simulated intensity patterns of the identified HOMO levels are superimposed at the bottom of each panel in Fig. 5b right below the experimental counterparts. Due to the remarkable agreement, we conclude that, after completion of the first saturated ML, subsequent NiTPP molecules adsorb in an undistorted, planar configuration. We note that for generating the theory maps, the azimuthal orientation of the planar molecules is slightly adjusted and the angle between the N-Ni-N axis and [001] direction is increased to 38°, as it improves the visual match between experimental and theoretical maps (see Fig. S10 in the SI for the 35° oriented maps in comparison). The fact that only molecules with explicit bonding to the Fe-O surface show traces of an altered macrocycle indicates that the saddle-shape is a consequence of the molecule-substrate interaction.

To extract information about how the planar conformation affects the energy splitting of HOMO and HOMO-1, we perform an analysis of the first peak in the EDC at around 1.8 eV. In doing so, we obtain a FWHM of (300 ± 100) meV for the peak associated with the second molecular adlayer (see. Sec. S6 for details on the fitting procedure). This peak width is considerably lower than the 500 meV we reported earlier for the first layer molecules. This reinforces the theoretical predictions regarding a relationship between conformation and the relative energy positions of HOMO and HOMO-1: for a selected MTPP we expect that a less (more) pronounced saddle-shape distortion leads to a reduced (increased) energy splitting of these orbitals.

Finally, we compare the energy position of molecular peaks for first and second layer NiTPP molecules, where we observe a shift of 300 meV. Taking into the account the small extent of this shift, we assign it to an attenuated surface screening, as done already[34], rather than distortion-induced variations.

**Conclusion:**

In summary, we investigated the interfaces formed by ZnTPP and NiTPP adsorbed on the oxygen-passivated Fe(100)-$p(1 \times 1)$O surface by employing the photoemission orbital tomography approach. In the saturated monolayer cases, the momentum-resolved data could confirm both the conformation and the azimuthal alignment of the molecules reported previously. In the case of ZnTPP, we were able to unambiguously identify the first double-structured peak as HOMO and HOMO-1, which are visibly separated in energy. While the reported minor saddle-shape conformation of these molecules has a noticeable influence on the momentum maps, their overall shape is akin to the one expected from planar molecules. On the contrary, the maps of the well-ordered 1 ML NiTPP film show distinct features that can be only linked to a strong saddle-shape distortion. When increasing the coverage of the NiTPP molecules to 2 ML, a clear LEED pattern, together with well-defined photoemission maps, confirms that the molecular order of the first layer prevails and is also transferred to the second layer. However, molecules in the first layer remain in their strong saddle-shape distorted state while the ones in the second layer show features of a planar structure indicating a decisive role of the molecule-substrate interaction with respect to molecular conformations. Furthermore, by interpreting measured peak widths with the help of DFT calculations, we find that a saddling of the macrocycle is accompanied by an increased energy splitting of HOMO and HOMO-1. An additional comparison between the 1 ML ZnTPP and 1 ML NiTPP experiments demonstrates that the actual extent of this splitting is crucially affected by the selection of the chelated metal ion; in our case, these metal ions differ mostly by the filling of their $d$ orbitals. While this filling does not change the appearance of the two highest MOs (HOMO and HOMO-1), it influences the BE of the HOMO, which is partially localized on the central metal ion. Our results demonstrate that POT can provide a full characterization of the molecule/metal interface, even for multilayer systems. From measurements performed at a single photon energy, it is possible to probe simultaneously the energy level alignment, the azimuthal orientation, and the adsorption geometry of complex aromatic molecules, such as metal-containing tetraphenyl porphyrins. Moreover, our data show that, while oxygen passivation is a viable way to decouple the organic molecule from the substrate, the interaction between the chelated ion and the surface is not completely suppressed. Crucially, the metal center plays a major role in determining the overall adsorption properties of the porphyrin, inducing, in the case of Ni, a prominent distortion of the macrocycle. This behavior is in contrast to previous reports that NiTPP has a planar macrocycle and is proven to be inert in multilayer films and on other oxygen-passivated surfaces, such as O-Cu(100). Our new findings suggest that the Fe-O interface represents a viable platform to study the reactivity of an electronically decoupled molecule in a distorted adsorption configuration.

**Acknowledgments:**

We acknowledge funding from the DFG (Major Research Instrumentation Individual Proposal INST 212/409-1), the Ministerium für Kultur und Wissenschaft des Landes Nordrhein-Westfalen (NRW). This paper was supported by EC H2020 programme under grant agreement No 965046, FET-Open project INTERFAST (Gated interfaces for fast information processing). The research leading to this result has been partially supported by the project CALIPSOplus under Grant Agreement 730872 from the EU Framework Programme for Research and Innovation HORIZON 2020. We further acknowledge funding from the European Research Council (ERC) under the European Union's Horizon 2020 research and innovation programme (Grant Agreement No. 725767—hyControl).

# Supplementary Information

## S1: The Fe(100)-p(1 × 1)O substrate

In Fig. S1, symmetrized momentum-resolved photoemission measurements for the pristine Fe(100)-p(1 × 1)O (Fe-O) surface, exhibiting the same characteristic features and bands well-known from literature, are shown.[1] The relation between the parallel momentum axes ($k_y$, $k_y$) and the substrate crystal directions ([001], [010]), which is illustrated in Fig. S1a, holds for all presented momentum maps throughout SI and main text.

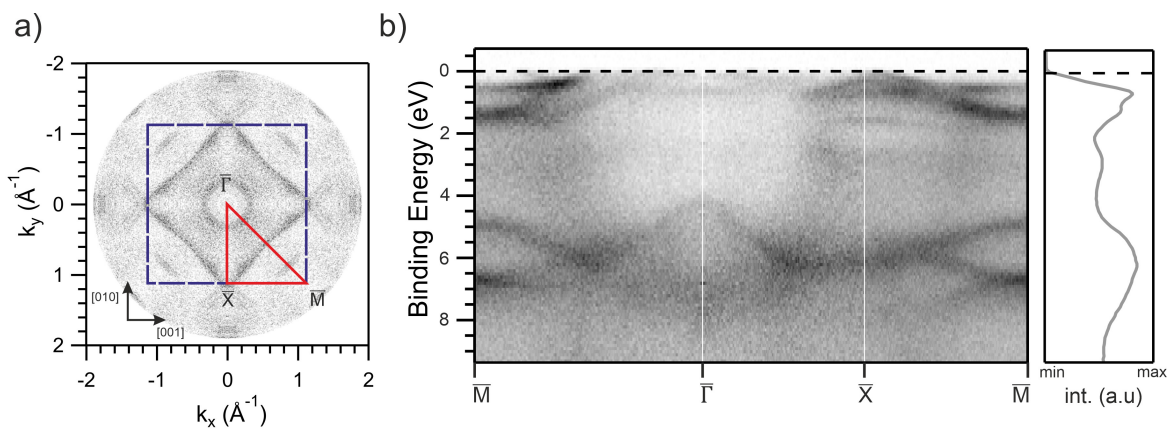

*Figure S1: Photoemission data of the pristine Fe(100)-p(1 × 1)O measured using an HHG laser source with a photon energy of 29.7 eV. In a) a symmetrized momentum map at the Fermi energy is shown, alongside the 1st surface BZ (violet square) and the corresponding high-symmetry lines (red). Cutting the 3D data cube along those lines yields the surface band structure, which is shown in b). The corresponding EDC of this spectrum is displayed on the right side of b).*

## S2: Adsorption geometries

In Fig. S2 we report basic models of the proposed adsorption orientations for ZnTPP and NiTPP molecules atop Fe-O that were created using the VESTA software package.[2] Adsorption site and azimuthal orientation of this schematic match the values proposed by Fratesi *et al*.[3] The angle between the [001] axis and the N-M-N axis (M = Zn or Ni) amounts to 17° for ZnTPP and 35° for NiTPP, respectively, as indicated by the solid black lines. For simplicity, no molecular distortions are taken into account.

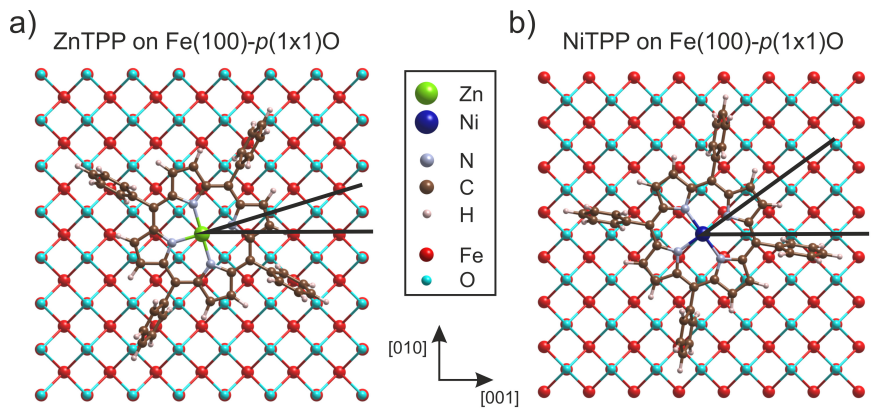

*Figure S2: Assumed adsorption geometries for a) ZnTPP and b) NiTPP on Fe(100)-p(1 × 1)O.*

## S3: Photoemission Orbital Tomography and Influence of the saddle shape

Photoemission orbital tomography (POT) offers the possibility to identify unambiguously molecular orbitals from photoemission experiments, by comparing the momentum distributions of the emitted photoelectrons to simulated momentum maps, which are extracted from the modulus square of the Fourier transform ($|FT|^2$) of individual orbitals.[4] An illustration of this concept is shown in Fig. S3 for the HOMO-1 orbital of ZnTPP.

a) real space orbital    b) $|FT|^2$    c) spherical cut    d) momentum map

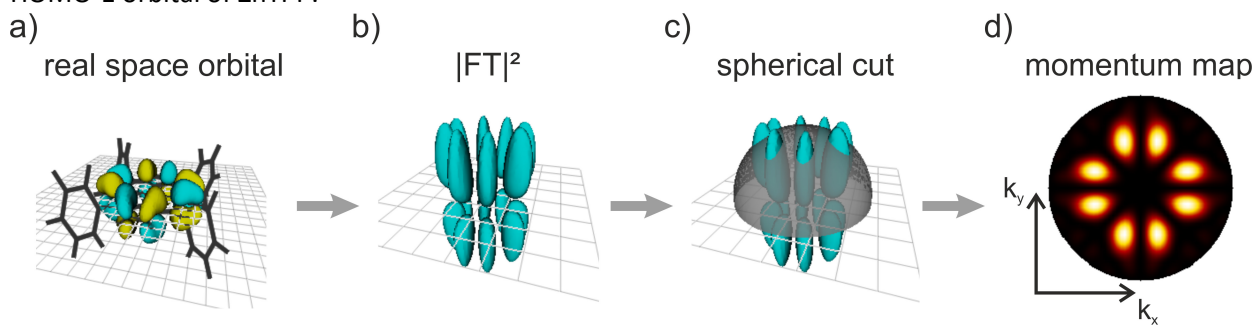

*Figure S3: Illustration of the photoemission orbital tomography approach. In a) the HOMO-1 orbital of a ZnTPP molecule is shown and in b) the corresponding $|FT|^2$. By cutting the $|FT|^2$ along the isoenergetic sphere indicated in c), the final momentum map, which is presented in d), is created.*

As shown by Hurdax et al.[5], the photoemission distribution of distorted organic molecules can have distinct fingerprints that deviate from the patterns of non-distorted molecules. In order to demonstrate how the molecular conformation alters the momentum maps, we consider the HOMO of ZnTPP for two different conformations and examine the resulting $|FTs|^2$. The real space distributions of the HOMO for both a planar and a saddle-shaped structure are depicted in Fig. S4a. Next to it, in Fig. S4b, the corresponding $|FTs|^2$ are displayed alongside selected isoenergetic spheres. The clear differences in the two $|FTs|^2$ indicate a significant influence on the ensuing momentum maps.

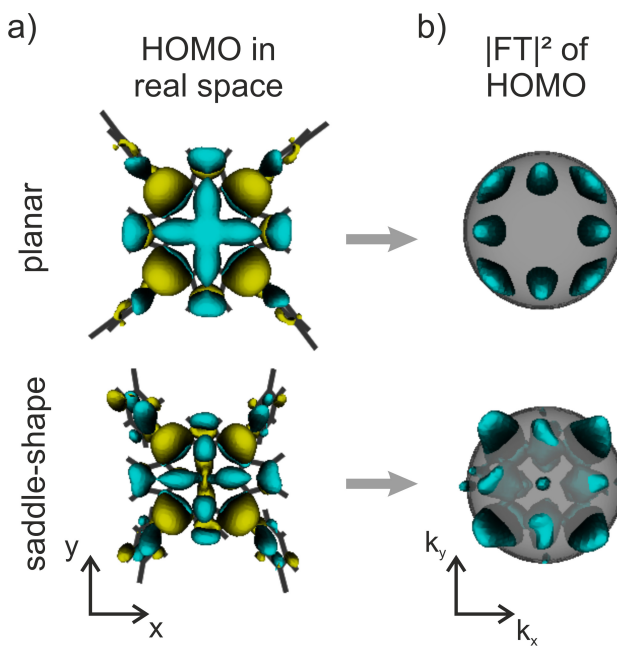

By taking into account the correct molecular adsorption geometry, simulated maps that serve as direct counterparts to the experimental momentum maps can be generated (see Fig. S10 for further details), thus, enabling an evaluation of the type and the degree of distortion *via* POT. In Fig. S5 the influence of different conformations is presented for the HOMO and HOMO-1 of ZnTPP. Upon an increasing saddle-shape, we observe the emergence of additional features around the center of the corresponding momentum maps (see Fig. S5a-c). Notably, the intensity variations induced by a ruffled structure (Fig. S5d) look considerably different compared to the saddle-shape induced ones; the ruffled geometry tends to be energetically favored for isolated porphyrins.[6] To simulate the presented maps, we chose a kinetic energy of 25 eV, which matches approximately the kinetic energy of the photoelectrons in our ZnTPP experiments. An analogous comparison for the simulated maps of NiTPP is shown in Fig. S6. Here, the assumed adsorption geometry corresponds to the one from Fig. S2b and the kinetic energy for extracting the |FT|² is set to 35 eV, accordingly to the used photon energy of 40 eV.[7] The maps in Fig. S6 consist of the sum of HOMO and HOMO-1, considering the prevalent degeneracy of these two orbitals. For extremely distorted conformations, a significant shift of the main photoemission intensity by 45° is observed, accompanied by rising intensities near the center of the map. Notably, a comparison of the same maps generated at kinetic energy of 25 eV (not shown here), demonstrates that the general appearance of the maps does not change upon changing the photon energy. However, at higher photon energies, the effects of saddling are more pronounced, suggesting an effective dependence on the photon energy as reported previously.[5] Thus, selecting the proper kinetic energy when generating the momentum map is crucial to ascertain the degree of saddling. Furthermore, by changing the experimentally used photon energy one may increase the accuracy of POT for detecting saddle-shape related features.

*Figure S4: influence of the saddle-shape on molecular orbitals. In a) the real space HOMO of ZnTPP is displayed for two different conformations (planar and saddle-shape). The corresponding $|FTs|^2$ are depicted in b).*

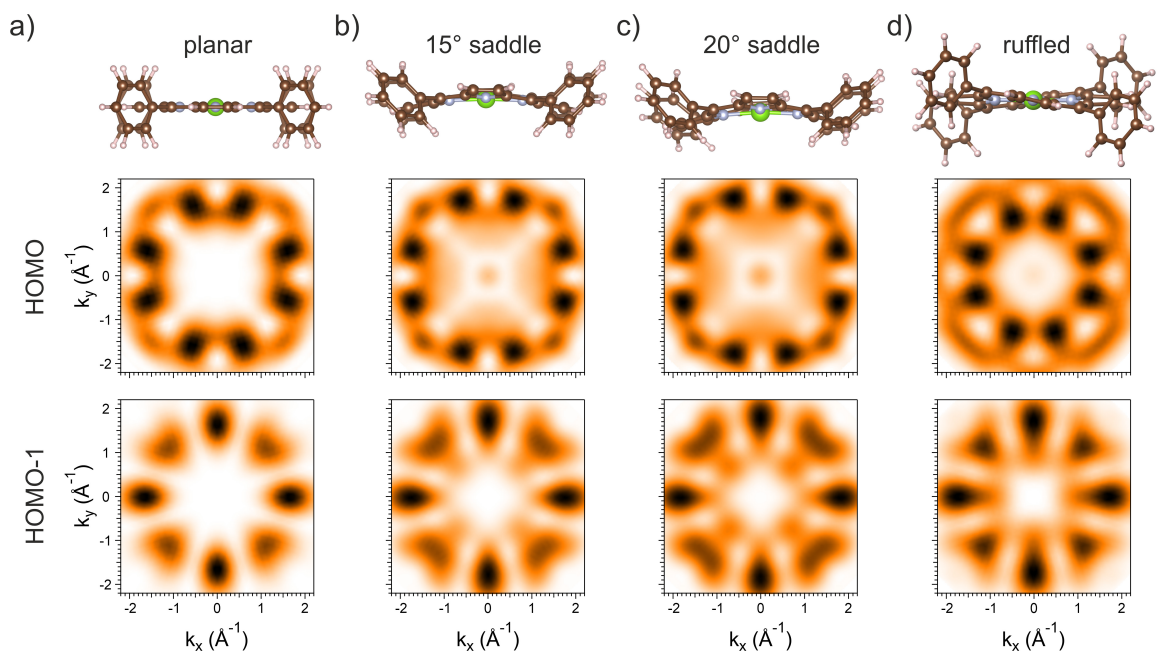

*Figure S5: Influence of molecular distortions on the HOMO and HOMO-1 momentum maps of ZnTPP. a)-d) Ball-and-stick models (top) of different macrocycle conformations and the corresponding simulated photoemission maps (bottom).*

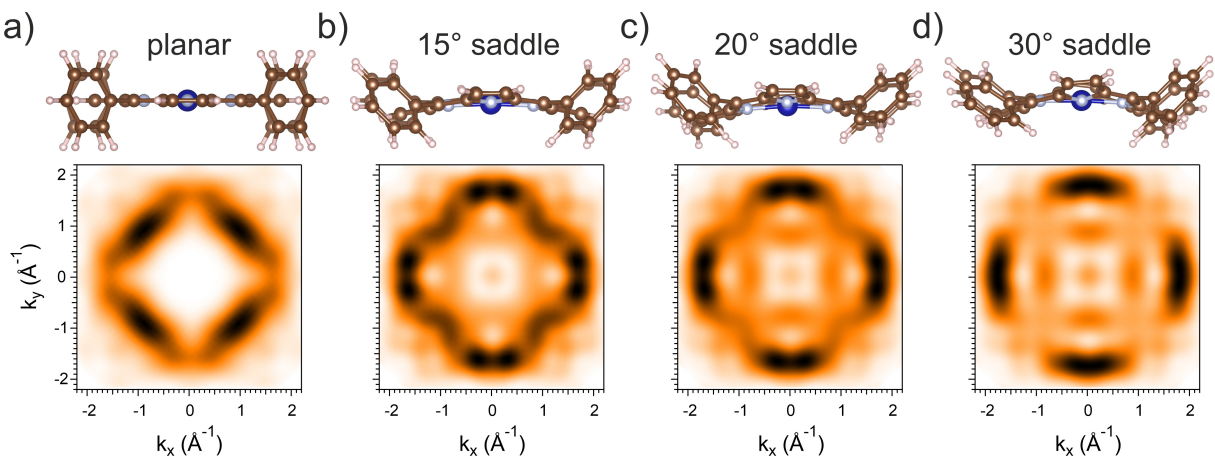

*Figure S6: Different saddle-shape conformations of NiTPP and their expected impact on the momentum space pattern. In a)-d) increasingly strong distortions of the macrocycle (top) and the corresponding simulated momentum maps of a superposition of HOMO and HOMO-1 (bottom) are shown.*

All the presented maps of this section were symmetrized according to the D$_4$ substrate surface symmetry.

### S4: Projected density of states

To emulate theoretically the EDCs we obtained *via* photoemission experiments, we calculate the density of states for the interface formed by 1ML NiTPP (and 1ML ZnTPP) adsorbed on Fe-O, in a geometrically optimized structure. For further insight into the occupied frontier orbital structure, we project the DOS on the total molecule and on the HOMO and HOMO-1 separately. The resulting plots for both ZnTPP/Fe-O and NiTPP/Fe-O are displayed in Fig. S7a and Fig. S7b, respectively. To match the experimental data, we introduce a Gaussian broadening of 350 meV, which is in accord with the FWHMs we determined for the peaks at the ZnTPP/Fe-O interface.

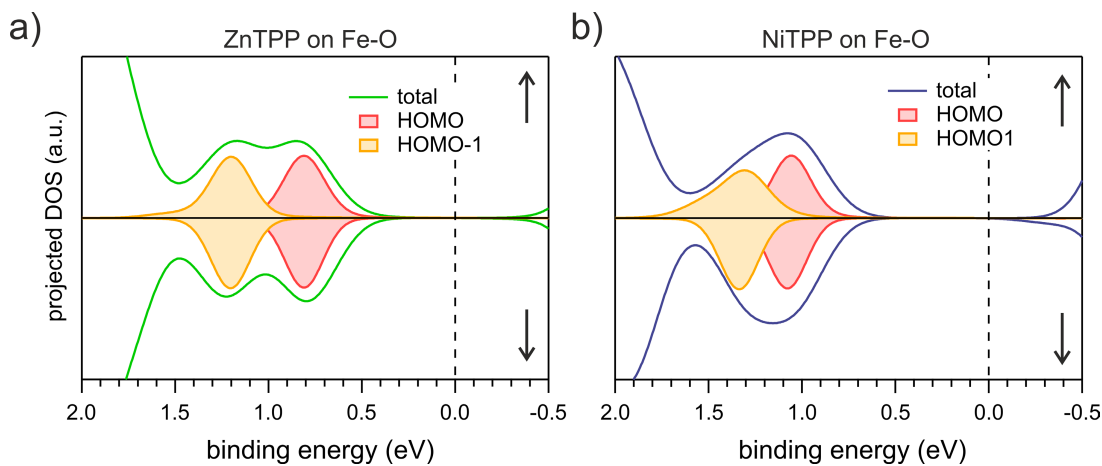

*Figure S7: Projected density of states for a) ZnTPP and b) NiTPP adsorbed on top of the Fe-O surface. The filled curves display the DOS projected on the HOMO (red) and HOMO-1 (orange). In addition, for both cases the DOS projected on the whole molecule is depicted by the green and blue curve for ZnTPP and NiTPP, respectively.*

## S5: Data analysis procedure

When using a PEEM, adjustments in the sample distance and in the microscope settings can have an influence in the size the field-of-view in momentum space, either increasing or decreasing it. Consequently, the scaling of the momentum maps for bare Fe-O and porphyrin-covered NiTPP/Fe-O can vary, which has to be considered before background subtraction. In order to ascertain the resizing factor, we compare photoemission maps at an energy where the features in *k*-space stem solely from the substrate (see Fig. S8a). By evaluating line profiles along a selected direction (dashed lines in the maps in Fig. S8a) we estimate the conversion factor and interpolate the substrate map dimensions accordingly. Afterwards the image is cut to the original pixel size yielding the final map presented in Fig. S8b.

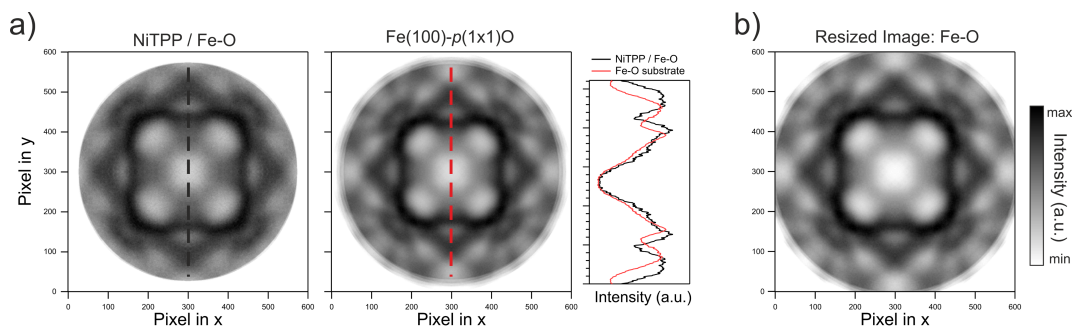

*Figure S8: Adjusting size of the momentum maps for before subtraction. a) Reference momentum maps of the NiTPP covered Fe-O film (left) vs. the bare Fe-O substrate (middle), which are integrated over the binding energy range of 0.5 eV to 0.8 eV. Dashed lines indicate the origins of the line profiles, which are presented in the right panel of a). To account for differently sized momentum maps the substrate map is interpolated according to the mismatch in the line profiles. The resized map is shown in b).*

Afterwards Fe-O maps of the same binding energy are subtracted from the NiTPP/Fe-O data. For the resized control map (normalization map), this procedure is shown in Fig. S9a. Since no molecular features are expected at this energy of 0.6 eV, the result is an almost homogenous pattern. In contrast, at a binding energy of 1.5 eV, subtraction gives rise to a distinct pattern that can be assigned to molecular features.

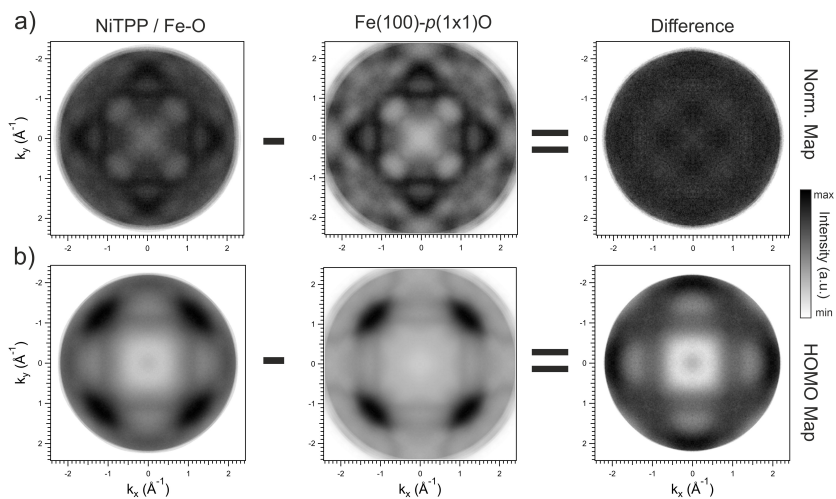

*Figure S9: Illustration of the subtraction procedure for the 1ML NiTPP film. In a) the background correction is shown for a map at 0.6 eV binding energy. This map does not include any molecular features. The left image shows the NiTPP/Fe-O sample, the middle one the Fe-O and the right one the result of the subtraction. In b) the same correction procedure is applied to the map at 1.5 eV, which corresponds to the momentum distribution that is shown in Fig. 3c.*

To compare the experimental data to the theoretical maps, the |FT|² cuts of the molecular orbitals are firstly oriented according to the assumed azimuthal geometry and, at the same time, for the degenerate orbitals the corresponding maps are summed up. For 1 ML NiTPP, the estimated angle between the N-Ni-N axis and the substrate [100] axis amounts to 35°, resulting in the HOMO/HOMO-1 image that is presented in Fig. S10a. After symmetrizing this map according to the $D_4$ substrate symmetry (to consider symmetry-equivalent domains), we end up with the map displayed in Fig. S10b. Additional symmetrized maps of the HOMO-2/HOMO-3 and HOMO-5 for this azimuthal orientation are summarized in Fig. S10c,d.

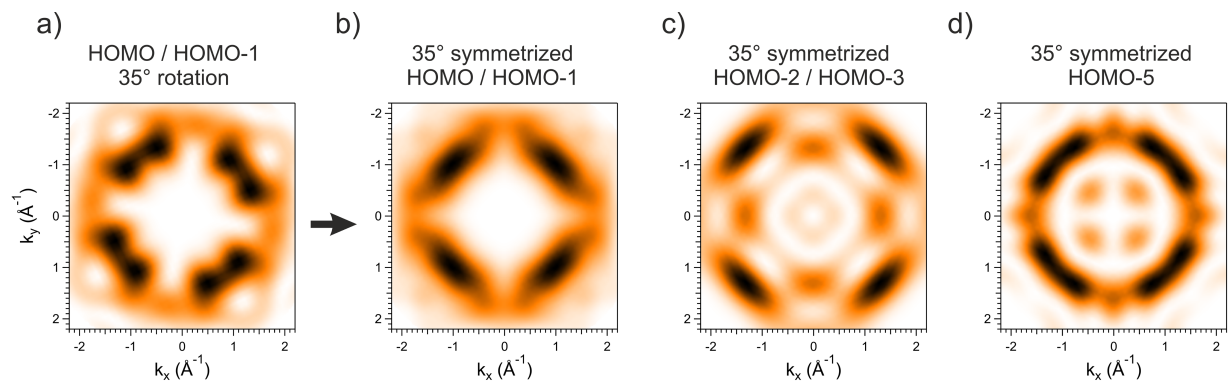

Figure S10: Generation of the simulated photoemission maps. In a) the averaged map of HOMO and HOMO-1 of planar NiTPP is displayed. This map corresponds to a molecular orientation of 35° between the molecule N-Ni-N axis and the susbtrate [001] direction. The resulting symmetrized maps of planar HOMO/HOMO-1, HOMO-2/HOMO-3 and HOMO-5 are displayed in b), c), d), respectively.

## S6: Evaluation of the HOMO/HOMO-1 peak for 2 ML NiTPP

As discussed in the main text, depositing a second layer of NiTPP molecules on top of the saturated 1 ML NiTPP/Fe-O film leads to the emergence of an additional resonance at around 1.7 eV. The corresponding EDC of the 2 ML NiTPP/Fe-O system is displayed in Fig. S11. Based on the appearance in our momentum maps, we can conclude that the resonance at 1.7 eV stems from the degenerate HOMO and HOMO-1 of planar molecules in the second organic adlayer. To gain insight into how planarity alters the relative energy positions of the two contributing orbitals, we extract the FWHM of this peak by a Gaussian fit like the one we performed on the 1 ML NiTPP sample. In doing so, we include a linear background and a second Gaussian, to also consider the signal originating from the first molecular layer at around 1.45 eV. The resulting curves are added to the plot in Fig. S11. Based on this fitting we estimate a FWHM of (300 ± 100) meV (indicated by the orange arrow). We note that for this particular peak, the applied fitting procedure is not fully accurate as the fit slightly deviates from the measured curve. Therefore, we estimate an error of 100 meV, which we also assume for all the other determined FWHMs.

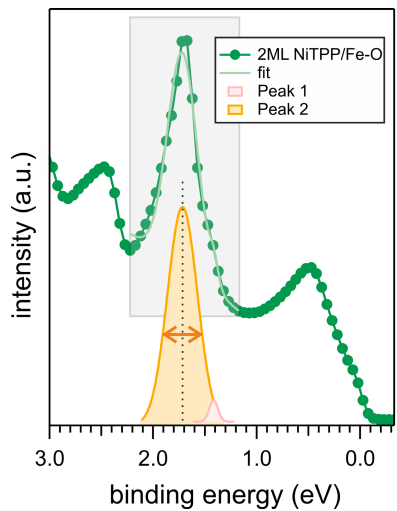

Figure S11: Illustration of the applied fit procedure for the 2 ML NiTPP/Fe-O data. The EDC (dark green curve) is fitted in the energy region highlighted by the grey box, using the sum of two Gaussians and a linear background as fitting function. For clarity, the resulting Gaussian contributions are added to the plot in form of the orange and the light red curves alongside the curve of the total fit (light green).